\documentclass[aps,twocolumn,superscriptaddress]{revtex4-2}

\usepackage[T1]{fontenc}
\usepackage{amssymb}
\usepackage{amsthm}
\usepackage{xfrac}
\usepackage{graphicx}
\usepackage{amsmath}
\usepackage{amsfonts}
\usepackage{color}
\usepackage{centernot}
\usepackage{mathtools}
\usepackage{stmaryrd}
\usepackage{xcolor}

\begin{document}

\title{Perturbative anomalous exponents from Kolmogorov multipliers}
\author{Alexei A. Mailybaev}
\affiliation{Instituto de Matem\'atica Pura e Aplicada -- IMPA, Rio de Janeiro, Brazil}
\email{alexei@impa.br}
\author{Simon Thalabard}
\affiliation{Institut de Physique de Nice, Universit\'e C\^ote d'Azur, Nice, France}
\email{simon.thalabard@univ-cotedazur.fr}
\begin{abstract}
Intermittency, manifested through anomalous scaling, remains one of the central unresolved problems in turbulence theory, with few analytical approaches extending beyond idealized linear transport models. We introduce a perturbative framework for anomalous scaling in turbulent transport based on multiplier statistics, rather than zero-mode calculations. We demonstrate the approach using a shell model combining deterministic and Kraichnan-like stochastic components. We reduce the problem to the analysis of a stationary Fokker--Planck equation for Kolmogorov multipliers, defined as ratios of successive scalar amplitudes. Its solution yields the invariant measure through a perturbative expansion around a Gaussian distribution. Using the resulting multiplier statistics, we compute explicit anomalous scaling exponents for structure functions of arbitrary order, including odd, even, and non-integer moments. Although demonstrated here for a shell model, the framework suggests a systematic perturbative route toward analytical theories of intermittency in turbulence.
\end{abstract}

\date{\today}
\maketitle

Intermittency---the anomalous nonlinear scaling of multipoint correlators---is a fundamental manifestation of fully developed turbulence, reflecting the breakdown of classical statistical scale invariance~\cite{frisch1995turbulence}. Explaining its origin has led to two broad lines of approach: cascade scenarios based on Kolmogorov multipliers~\cite{kolmogorov1962refinement,benzi1993intermittency,eyink2003gibbsian} and analytical descriptions based on zero modes~\cite{bernard1998slow,falkovich2001particles,benzi2023lectures}. 
While both approaches have achieved important successes, neither has yet led to a complete predictive framework.

By postulating that turbulent advection induces a multiplicative stochastic process across scales, cascade scenarios provide statistical descriptions of intermittency, ranging from the pioneering Kolmogorov--Obukhov lognormal model~\cite{kolmogorov1962refinement} to more sophisticated approaches, many of which are consistent with experimental and numerical observations~\cite{benzi1993intermittency,dubrulle1994intermittency,eyink2003gibbsian,chen2003kolmogorov,ruffenach2026spatio}. For example, the time-dependent random multiplicative process framework~\cite{benzi2003intermittency} computes anomalous exponents using a stochastic closure based on multiplier dynamics. A formal connection between such phenomenological descriptions and the equations of motion is provided by the hidden-symmetry approach~\cite{mailybaev2021hidden,mailybaev2022hiddenMT}, which posits a weak form of statistical scale invariance underlying turbulent intermittency and relates anomalous exponents to Perron--Frobenius eigenvalues associated with scale dynamics~\cite{mailybaev2022hidden,mailybaev2023hidden}. 
While numerical evidence for hidden symmetry has been reported in various transport models~\cite{thalabard2024zero,magacho2025scale,calascibetta2025hidden}, the theory is not yet fully predictive. In particular, analytical predictions for anomalous exponents remain unavailable, except in certain solvable constructions~\cite{mailybaev2021solvable}.

On the other hand, zero-mode theory has emerged as a paradigmatic framework owing to its analytical success in the related problem of Kraichnan scalar transport~\cite{falkovich2001particles,cardy2008non}. In that setting, a passive scalar is linearly advected by a Gaussian random velocity field with white-in-time statistics and monofractal spatial scaling exponent $\xi$. The scalar correlators satisfy a closed Hopf hierarchy at each order, leading to anomalous exponents in the form of zero-mode solutions. These exponents can be computed perturbatively by exploiting the Gaussian structure in the rough limit ($\xi\to0$)~\cite{gawedzki1995anomalous,vergassola1997structures}, the smooth limit ($\xi\to2$)~\cite{pumir1997perturbation,shraiman2000scalar}, or the large-dimensional limit~\cite{chertkov1996anomalous}. 
The main limitation of the approach is the difficulty of extending zero-mode calculations to nonlinear settings, where the Hopf hierarchy is no longer closed~\cite{arad2001statistical,angheluta2006anomalous}.

In this Letter, we show that anomalous exponents can be computed perturbatively by bypassing the Hopf hierarchy and working directly with Kolmogorov multipliers. To this end, we introduce a random shell model allowing for a perturbative expansion around a Kolmogorov fixed point. 
The problem reduces to analyzing a stationary Fokker--Planck equation for multiplier statistics, whose invariant measure is constructed as a perturbative expansion around a Gaussian distribution. 
The resulting multiplier statistics yield explicit anomalous exponents for structure functions of arbitrary order, including odd, even, and non-integer moments.

\textit{Model.}
Random shell models for passive scalar advection~\cite{wirth1996anomalous,benzi1997analytic,andersen1999shell} follow the standard shell-model framework for turbulence~\cite{frisch1995turbulence,biferale2003shell}, where multiscale dynamics is discretized in wavenumber space as $k_n = \lambda^n$ with $\lambda>1$ (typically $\lambda=2$). Scalar increments are represented by real variables $\theta_n(t)$ associated with shells $n\ge1$. A minimal model reads~\cite{jensen1992shell,benzi1997analytic,biferale2007minimal}
\begin{equation}
\left(\frac{d}{dt}+\kappa k_n^2\right)\theta_n
= k_{n-1}\theta_{n-1}u_{n-1}-k_n \theta_{n+1}u_n,
\label{eq1}
\end{equation}
where $\kappa\ge0$ is the diffusivity. The nearest-neighbor interactions conserve the scalar energy $E_\theta=\sum_n\theta_n^2$ in the inviscid case $\kappa=0$. Large-scale forcing is imposed by setting $\theta_0=1$.

Following the ideas of Kraichnan~\cite{kraichnan1968small,falkovich2001particles,frisch2007intermittency,cardy2008non}, we model the advecting velocities as
\begin{equation}
u_n(t) = k_n^{-1/3}+\varepsilon k_n^{-2/3} \xi_n(t),
\label{eq2}
\end{equation}
where $\xi_n(t)$ are independent white noises.
The velocity field~\eqref{eq2} is chosen to satisfy the space--time scale invariance of Kolmogorov (K41) theory,
\begin{equation}
u_n(t) \mapsto \lambda^{1/3} u_{n+1}(\lambda^{-2/3}t).
\end{equation}
It consists of a deterministic component corresponding to the K41 self-similar state and a stochastic component with amplitude $\varepsilon$. The latter serves as the perturbation parameter for a systematic expansion around the K41 state, in the same spirit as perturbative expansions in the Kraichnan model.
The products in Eq.~\eqref{eq1} are understood in the Stratonovich sense. 

\begin{figure}[tp]
\centering
\includegraphics[width=0.45\textwidth]{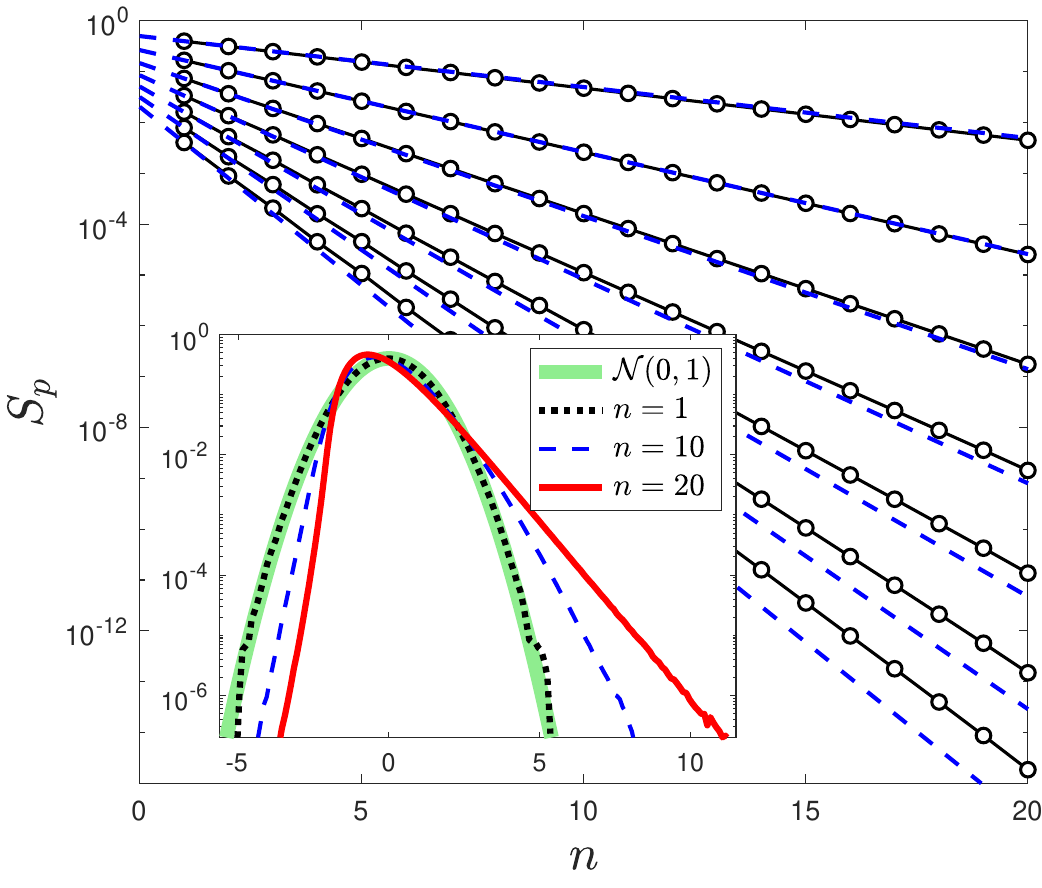}
\caption{Power-law scaling of the structure functions $S_p(n)$ in the inertial interval for $\varepsilon = 0.1$ and $p=1,\ldots,7$. Solid lines with circles denote results from numerical simulations, while dashed lines show the Obukhov--Corrsin scaling $\propto k_n^{-p/3}$. The lines are vertically shifted for clarity. Inset: PDFs of the normalized shell variables $(\theta_n-k_n^{-1/3})/\left[\mathrm{Var}(\theta_n)\right]^{1/2}$ for $n=1,10,20$, compared with the Gaussian distribution.}
\label{fig1}
\end{figure}

For $\kappa \ll 1$, the system exhibits a forcing range at small wavenumbers \(k_n\sim1\) and a diffusion range at large \(k_n\), separated by an inertial interval characterized by intermittent dynamics. In this inertial interval, the structure functions $S_p(n)=\langle|\theta_n|^p\rangle \propto k_n^{-\zeta_p}$ display power-law scaling with anomalous exponents $\zeta_p$; see Fig.~\ref{fig1} and Supplemental Material (SM)~\cite[\S VI]{SM} for details of the numerical simulations. Deviations of these exponents from the Obukhov--Corrsin scaling $\propto k_n^{-p/3}$ provide a clear signature of intermittency in turbulence~\cite{frisch1995turbulence}. In the inset of Fig.~\ref{fig1}, we also show the normalized PDFs of the shell variables $\theta_1$, $\theta_{10}$, and $\theta_{20}$ within the inertial interval. These PDFs exhibit increasingly pronounced non-Gaussian behavior toward smaller scales, reflecting the intermittent nature of the dynamics. The distributions are asymmetric due to their nonzero mean values and develop heavy tails toward large positive values of $\theta_n$, indicating an enhanced probability of strong fluctuations.

The main result of this paper is a first-principles analytical derivation of the expansion
\begin{equation}
\zeta_p
= \frac{p}{3}-\frac{\lambda^{2/3}+1}{4\lambda^{1/3} \ln\lambda}\,
p(p-2)\,\varepsilon^2
+O(\varepsilon^4),
\label{eq_zeta_fin}
\end{equation}
which provides   a quadratic nonlinearity  for the anomalous exponents of the structure functions for arbitrary order $p \in \mathbb{R}$. The derivation is based on an $\varepsilon$-expansion of the joint distribution of Kolmogorov multipliers, 
which is consistent with the restoration of hidden symmetry of the equations of motion~\cite{thalabard2024zero}.

\textit{Multipliers and hidden symmetry.}
Neglecting diffusion, Eqs.~(\ref{eq1})--(\ref{eq2}) can be written in the inertial interval as the SDE system
\begin{equation}\label{eq:strat-theta-app}
\begin{aligned}
d\theta_n
= \,&
\Bigl(
\gamma^{2n-2}\theta_{n-1}
-\gamma^{2n}\theta_{n+1}
\Bigr)\,dt
\\&
+\varepsilon 
\Bigl(
\gamma^{n-1}\theta_{n-1}\circ dw_{n-1}
-\gamma^n \theta_{n+1}\circ dw_n
\Bigr),
\end{aligned}
\end{equation}
where $\gamma=\lambda^{1/3}$ and $w_n(t)$ are independent Wiener processes. 
System~(\ref{eq:strat-theta-app}) is invariant under the scalings
\begin{equation}
\theta_n(t) \mapsto \alpha\,\theta_{n+s}(\gamma^{-2s}t),
\quad
w_n(t) \mapsto \gamma^s\,w_{n+s}(\gamma^{-2s}t),
\label{eq_sym}
\end{equation}
with arbitrary $s\in\mathbb{Z}$ and $\alpha\in\mathbb{R}$. For $\varepsilon=0$, this system admits the stationary Kolmogorov solution $\theta_n=\gamma^{-n}=k_n^{-1/3}$, which is invariant under the transformation~(\ref{eq_sym}) for the particular choice $\alpha=\gamma^s=\lambda^{s/3}$. However, intermittency at $\varepsilon>0$ breaks all symmetries~(\ref{eq_sym}) in the inertial interval.

Taking into account that $\theta_n/\theta_{n-1}=\gamma^{-1}$ for the Kolmogorov solution, we introduce the multipliers
\begin{equation}
x_n=\frac{\theta_n}{\theta_{n-1}}=\gamma^{-1}+\varepsilon z_n.
\label{eq_mult}
\end{equation}
We will be interested in the variables $z_n$, for which equation~(\ref{eq:strat-theta-app}) yields a closed SDE system
\begin{equation}
dz_n = a_n(z,\varepsilon)\,dt+\sum_{m} B_{nm}(z,\varepsilon)\,dw_m,
\label{eq_SDE_z}
\end{equation}
see SM~\cite[\S I]{SM} for the explicit expressions. 
In these variables, the symmetry~(\ref{eq_sym}) reduces to
\begin{equation}
z_n(t) \mapsto z_{n+s}(\gamma^{-2s}t),
\ \
w_n(t) \mapsto \gamma^s\,w_{n+s}(\gamma^{-2s}t).
\label{eq_HS}
\end{equation}
Notice that the arbitrary factor $\alpha$ is eliminated! Relations~(\ref{eq_HS}) are called the \emph{hidden symmetry}, which is an exact symmetry of the multiplier equations. It was previously supported numerically that this symmetry is restored asymptotically in the statistical sense within the inertial interval, even though the original symmetries~(\ref{eq_sym}) are all broken~\cite{thalabard2024zero}. Throughout this work, we adopt this hypothesis, which determines the functional form of stationary probability distributions, namely, invariance under translations $z_n\mapsto z_{n+s}$. The perturbative Fokker--Planck equations are then solved exactly at each perturbative order.

\textit{Distribution of multipliers.}
At $\varepsilon=0$, Eq.~(\ref{eq_SDE_z}) reduces to the Ornstein--Uhlenbeck process
\begin{equation}
dz_n = \sum_m A_{nm}^{(0)} z_m\,dt + \sum_m B_{nm}^{(0)}\,dw_m,
\end{equation}
with matrices
\begin{align}
A_{nm}^{(0)}
= & \gamma^{2n-3}\bigl(
\delta_{m,n-1}+(1-\gamma^2)\delta_{m,n}-\gamma^2\delta_{m,n+1}
\bigr), \\
B_{nm}^{(0)}
= &
-\gamma^{n-2}\delta_{m,n-2}
+\bigl(\gamma^{n-1}+\gamma^{n-3}\bigr)\delta_{m,n-1} 
\nonumber \\&
-\gamma^{n-2}\delta_{m,n}.
\label{eq_SDE_aBM}
\end{align}
Its stationary density is Gaussian,
\begin{equation}
\label{eq:Gaussian}
p_0(z)=Z^{-1}\exp\!\left(-\tfrac12 z^T C^{-1} z\right),
\end{equation}
where the covariance matrix $C$ satisfies the Lyapunov equation
\begin{equation}
\label{eq_for_C_inf}
A^{(0)}C+CA^{(0)T}+D^{(0)}=0, \quad D^{(0)}=B^{(0)}B^{(0)T}.
\end{equation}

\begin{figure}[tp]
\centering
\includegraphics[width=0.45\textwidth]{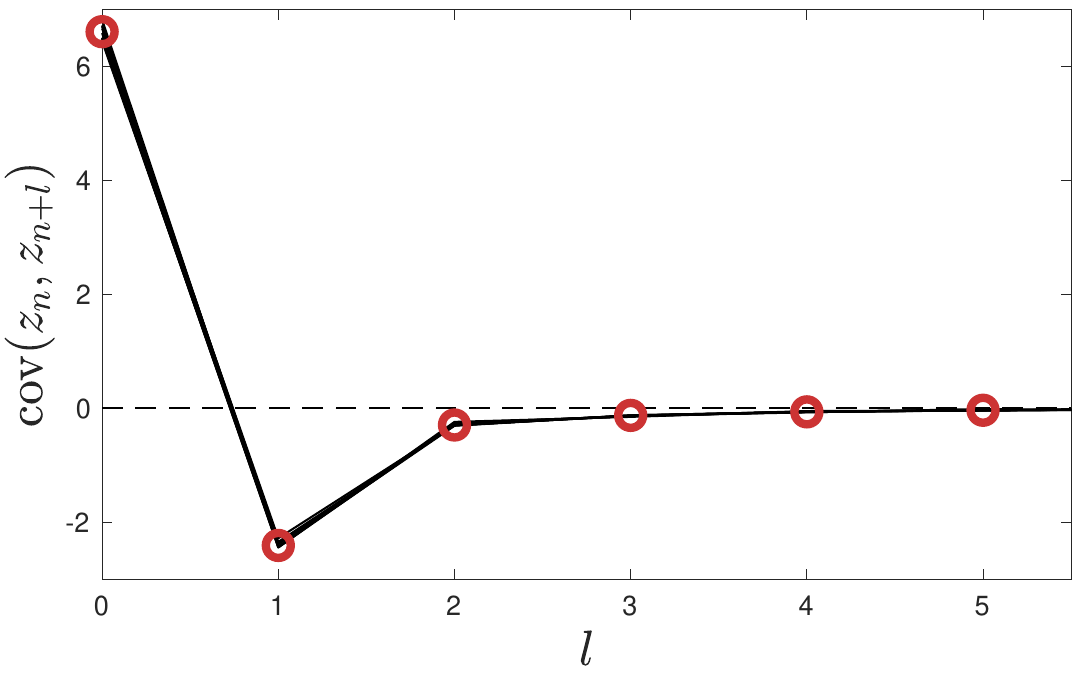}
\caption{Covariances $\mathrm{cov}(z_n,z_{n+l})$ from simulations at $\varepsilon=0.01$ (black lines for inertial-interval shells $n=6,\ldots,14$), compared with theoretical values $c_l$ (red circles).}
\label{fig2}
\end{figure}

\begin{figure}[tp]
\centering
\includegraphics[width=0.4\textwidth]{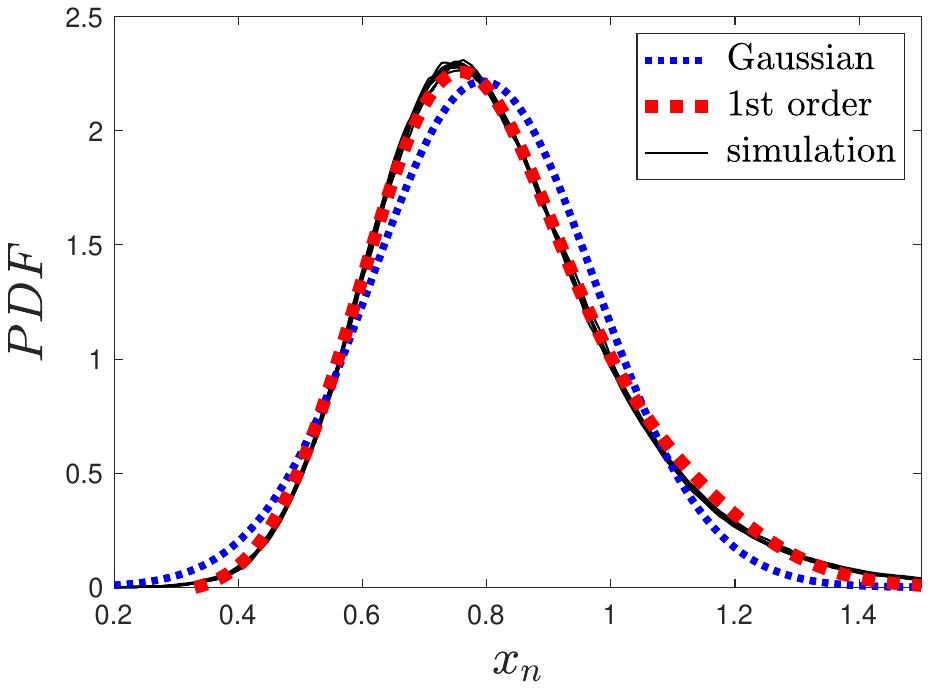}
\caption{Probability density functions (PDFs) of the multipliers $x_n=\gamma^{-1}+\varepsilon z_n$ for $\varepsilon=0.07$. Simulation results (black lines for inertial-interval shells $n=6,\ldots,14$) are compared with theoretical predictions: the Gaussian approximation (blue dotted line) and the higher-order non-Gaussian correction (red dotted line).}
\label{fig3}
\end{figure}

Assuming statistical restoration of the hidden symmetry in the inertial interval, the stationary distribution (\ref{eq:Gaussian}) must be invariant under shifts $n\mapsto n+s$, implying that the covariance matrix $C$ is symmetric Toeplitz, 
\begin{equation}
\label{eq_C_Toeplitz}
C_{nm}=c_{|n-m|}. 
\end{equation}
Equation~(\ref{eq_for_C_inf}) then reduces to an explicit linear system for $c_l$ as (see SM~\cite[\S II]{SM})
\begin{align}
2(1-\gamma^2)(c_0+c_1)+\gamma^{-3}(\gamma^4+4\gamma^2+1) &\, = 0,
\label{eq:c_l0M}
\\[2mm]
(1-\gamma^4)(c_1+c_2)-2(\gamma+\gamma^{-1}) &\, = 0,
\label{eq:c_l1M}
\\[2mm]
(1-\gamma^6)c_3+(1-\gamma^2)(1+\gamma^4)c_2 & \nonumber \\
+\,(\gamma^4-\gamma^2)c_1+\gamma &\, = 0,
\label{eq:c_l2M}
\\[2mm]
(1-\gamma^{2l+2})\,c_{l+1}
+(1-\gamma^2)(1+\gamma^{2l})\,c_l
& \nonumber
\\+\,(\gamma^{2l}-\gamma^2)\,c_{l-1}
= 0, \ \ l &\, \ge 3,
\label{eq:c_l3M}
\end{align}
which defines unique covariance coefficients under the condition $c_l \to 0$ as $l \to \infty$. Figure~\ref{fig2} shows the coefficients \(c_l\) for \(\gamma = 2^{1/3}\), compared with the covariances \(\mathrm{cov}(z_n,z_{n+l})\) obtained from numerical simulations.
For marginal distributions, the Gaussian law (\ref{eq:Gaussian}) implies $z_n\sim\mathcal{N}(0,c_0)$, also in agreement with numerical simulations; see Fig.~\ref{fig3}.


Higher-order corrections are obtained by expanding
\begin{equation}
\begin{aligned}
a(z,\varepsilon)=\, & a^{(0)}(z)+\varepsilon a^{(1)}(z)+\cdots,\\
B(z,\varepsilon)=\, & B^{(0)}+\varepsilon B^{(1)}(z)+\cdots,
\end{aligned}
\end{equation}
and writing
\begin{equation}
p(z,\varepsilon)=p_0(z)\bigl[1+\varepsilon r_1(z)+\varepsilon^2 r_2(z)+\cdots\bigr],
\label{eq_p_expansion}
\end{equation}
with the normalization conditions $\int r_k(z)p_0(z)\,dz=0$. Equations for the functions $r_k(z)$ follow from the $\varepsilon$-expansion of the stationary Fokker--Planck equation, which can be solved under the assumption of restored hidden symmetry. At first order, we obtain (see SM~\cite[\S III]{SM})
\begin{equation}
r_1(z)=m\sum_n q_n+\sum_{i,j,k} w_{ijk}\, {:}\,q_iq_jq_k\,{:},
\label{eq:r_all}
\end{equation}
where $q=C^{-1}z$ are dual variables, the coefficient
\begin{equation}
m=\frac{\gamma}{2}\,c_0-\frac{\gamma^2+1}{2\gamma^2},
\label{eq:m_final}
\end{equation}
and ${:}\,q_iq_jq_k\,{:}$ denotes Wick ordered product with respect to $p_0$.
The hidden symmetry \eqref{eq_HS} implies $w_{ijk}=W_{j-i,k-i}$, where the matrix $W$ is determined by the explicit system of linear equations given in SM~\cite[\S IIIA]{SM}.
For the marginal distribution, Eqs.~(\ref{eq:Gaussian}) and (\ref{eq:r_all}) yield (see SM~\cite[\S IV]{SM})
\begin{equation}
\begin{aligned}
p(z_n) =\, &
\frac{1}{\sqrt{2\pi c_0}}
\exp\!\left(-\frac{(z_n-\varepsilon m)^2}{2c_0}\right) 
\\
& \times\left[
1+\varepsilon \frac{W_{00}}{c_0^3}
\left(z_n^3-3c_0 z_n\right)
+O(\varepsilon^2)
\right].
\end{aligned}
\label{eq:marginal_shifted_final_simplified}
\end{equation}

For \(\gamma=2^{1/3}\), we obtain \(c_0\approx 6.6085\), \(m\approx 3.3481\), and \(W_{00}\approx 25.8962\). Figure~\ref{fig3} demonstrates excellent agreement between the theoretical prediction (\ref{eq:marginal_shifted_final_simplified}) and numerical simulations for \(\varepsilon=0.07\), where the non-Gaussian corrections are clearly visible.

\textit{Structure functions.}
Consider the structure functions \(S_p(n)=\langle|\theta_n|^p\rangle\). For small \(\varepsilon\), sign changes of the multipliers \(x_k=\gamma^{-1}+\varepsilon z_k\) have exponentially small probability and therefore do not affect the perturbative calculation below. Using the telescopic representation
\begin{equation}
\theta_n
= \prod_{k=1}^n x_k
= \gamma^{-n} \prod_{k=1}^n (1+\varepsilon \gamma z_k),
\end{equation}
we expand \(|\theta_n|^p\) in \(\varepsilon\) to obtain
\begin{equation}
\begin{aligned}
&S_p(n)
= 
\gamma^{-pn}
\bigg[
1
+p\varepsilon\gamma\sum_{k=1}^n\langle z_k\rangle
\\
&\ \ +\frac{\varepsilon^2\gamma^2}{2}
\bigg(
p^2\Big\langle\Big(\sum_{k=1}^n z_k\Big)^2\Big\rangle
-p\sum_{k=1}^n\langle z_k^2\rangle
\bigg)
+O(\varepsilon^3)
\bigg].
\end{aligned}
\label{eq:Sp_prelim}
\end{equation}
The averages can be evaluated using Wick's theorem together with the perturbative distribution~\eqref{eq_p_expansion}, including the Gaussian contribution~\eqref{eq:Gaussian} and the correction~\eqref{eq:r_all}. For large \(n\), this yields (see SM~\cite[\S V]{SM})
\begin{equation}
S_p(n)\propto \gamma^{-pn}\exp\left[
\frac{\gamma^2+1}{4\gamma}p(p-2)\,\varepsilon^2 n
+O(\varepsilon^3 n)\right].
\label{eq_Sp_1}
\end{equation}
Relation (\ref{eq_Sp_1}) can be written as the scaling law \(S_p(n)\propto k_n^{-\zeta_p} = \gamma^{-3 \zeta_p n}\) with the exponent 
\begin{equation}
\zeta_p
= \frac{p}{3}-\frac{\gamma^2+1}{12\gamma\ln\gamma}\,
p(p-2)\,\varepsilon^2
+O(\varepsilon^4),
\label{eq_zeta_fin_gamma}
\end{equation}
where the odd-order corrections vanish due to the symmetry \(\varepsilon\mapsto-\varepsilon\) in the statistics of system~(\ref{eq:strat-theta-app}).

Recalling that $\gamma = \lambda^{1/3}$ yields the final Eq.~(\ref{eq_zeta_fin}).  Here the zero-order term $p/3$ is the exponent of the Kolmogorov scaling corresponding to the self-similar solution $\theta_n = k_n^{-1/3}$.
The next term is proportional to $\varepsilon^2$ and determines the intermittency corrections.
Remarkably, the leading anomalous corrections for arbitrary order $p \in \mathbb{R}$ are expressed explicitly in terms of the parameter $\gamma$. Although we focus here on positive orders, the results extend to negative-order moments, provided that these remain finite in the inviscid limit.

\begin{figure*}[tp]
\centering
\includegraphics[width=0.95\textwidth]{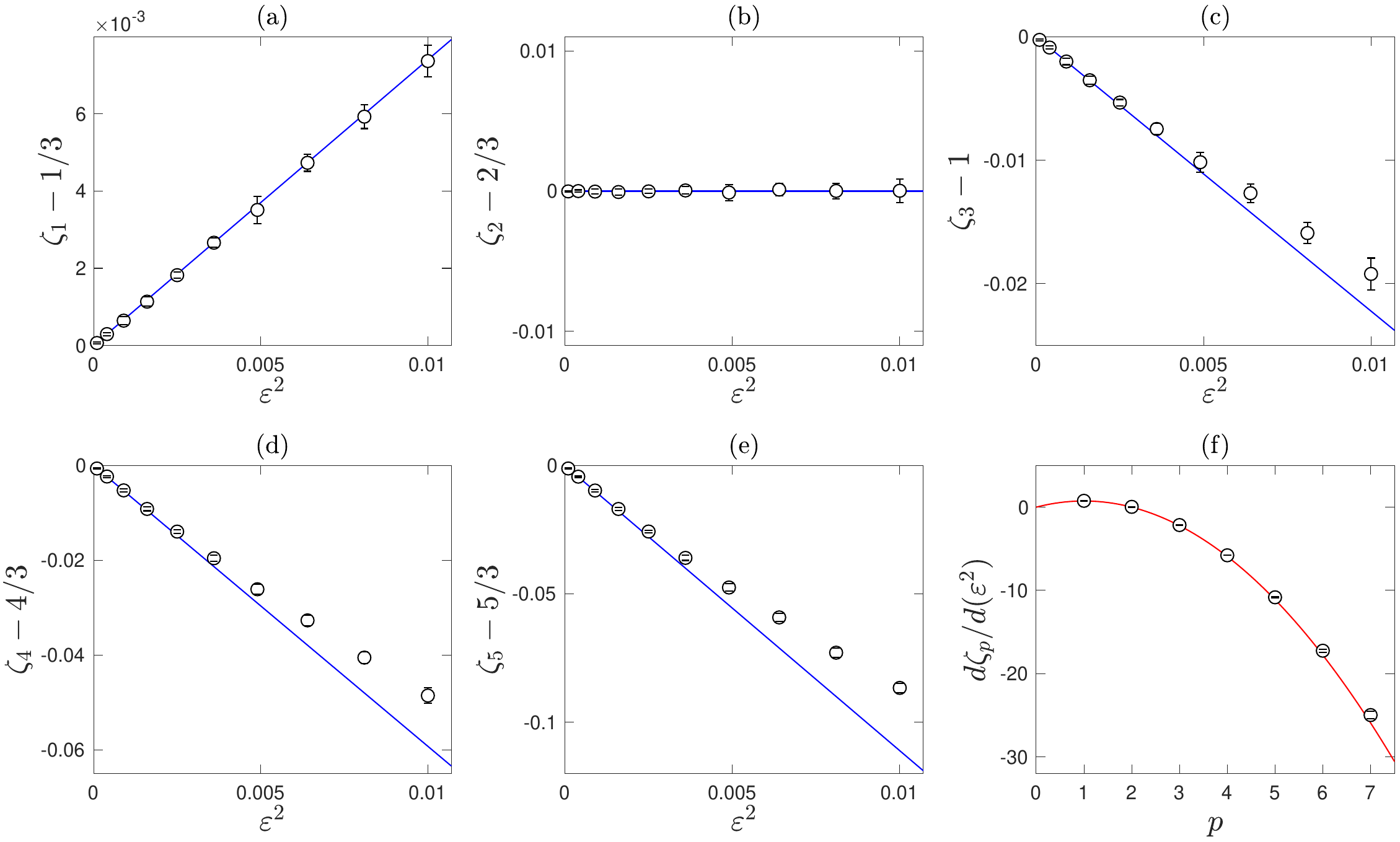}
\caption{(a--e) Anomalous corrections $\zeta_p-p/3$ as functions of $\varepsilon^2$ for $p=1,\ldots,5$. Solid lines show theoretical prediction~(\ref{eq_zeta_fin}); circles with error bars are results of numerical simulations for $\varepsilon = 0.01,0.02,\ldots,0.1$. (f) Slope $d\zeta_p/d(\varepsilon^2)$ at $\varepsilon=0$ as a function of $p$: theory (solid line) vs.\ numerical simulation estimates (circles).}
\label{fig4}
\end{figure*}

Figures~\ref{fig4}(a--e) show the anomalous corrections \(\zeta_p-p/3\) as functions of \(\varepsilon^2\) for \(p=1,\ldots,5\). The first-order predictions~(\ref{eq_zeta_fin}) (solid lines) are compared with numerical simulation results (circles), demonstrating the expected linear dependence at small \(\varepsilon^2\) and deviations at larger values. 
Figure~\ref{fig4}(f) shows the theoretical slope \(d\zeta_p/d(\varepsilon^2)\vert_{\varepsilon=0}\) predicted by Eq.~(\ref{eq_zeta_fin}) (solid line). This prediction is verified by extracting \(d\zeta_p/d(\varepsilon^2)\) from polynomial fits to the numerical data \(\zeta_p(\varepsilon^2)\) shown in Figs.~\ref{fig4}(a--e) (circles), demonstrating excellent agreement between theory and simulations.

\textit{Zero modes.}
For comparison with the perturbative theory developed above, we briefly discuss an alternative derivation of anomalous exponents based on zero modes. This approach relies on (and is therefore restricted to) the linear structure of the ideal model~(\ref{eq:strat-theta-app}) and is limited to integer values of \(p\). Moreover, the computation must be carried out separately for each \(p\), with complexity increasing rapidly as \(p\) increases.

The linearity of the model implies that the Hopf hierarchy closes at each order, allowing one to determine power-law zero modes directly. We illustrate this approach for the simplest case of the first-order moment.
The stationary Hopf equation for the mean values reads
\begin{equation}
0 =
\gamma^{2n-2}M_{n-1}
-\gamma^{2n}M_{n+1}
-\frac{\varepsilon^2}{2}
\left(\gamma^{2n-2}+\gamma^{2n}\right)M_n,
\label{eq:mean_hopf}
\end{equation}
where \(M_n=\langle \theta_n \rangle\).
Seeking a power-law solution of the form \(M_n = \gamma^{-3\zeta_1 n}\), we obtain
\[
\gamma^{3\zeta_1}
-\gamma^{2-3\zeta_1}
=
\frac{\varepsilon^2}{2}(1+\gamma^2),
\]
which yields
\begin{equation}
\zeta_1
=
\frac{1}{3\ln\gamma}
\ln\left(
\frac{\varepsilon^2}{4}(1+\gamma^2)
+
\sqrt{
\gamma^2+\frac{\varepsilon^4}{16}(1+\gamma^2)^2
}
\right).
\label{eq:zeta1_exact_log}
\end{equation}
For small \(\varepsilon\), expanding Eq.~\eqref{eq:zeta1_exact_log} gives
\begin{equation}
\zeta_1
=
\frac{1}{3}+\frac{\gamma^2+1}{12\gamma\ln\gamma}\,\varepsilon^2
+O(\varepsilon^4),
\label{eq:zeta1_expansion}
\end{equation}
in agreement with the general perturbative result~(\ref{eq_zeta_fin_gamma}).

\textit{Conclusion.}
In this work, we developed a perturbative framework for anomalous scaling based directly on multiplier statistics. Applied to a random shell model of turbulent transport, the approach yields explicit anomalous exponents for structure functions of arbitrary order together with a perturbative construction of the invariant multiplier statistics. At leading order, the model yields approximately Gaussian multiplier statistics consistent with lognormal phenomenology, while higher perturbative orders capture non-Gaussian corrections. The perturbative construction is systematic and can, in principle, be continued to arbitrary perturbative order.

The main conceptual point is that the framework is formulated directly in terms of the stochastic dynamics of Kolmogorov multipliers. Unlike zero-mode approaches, it does not rely on the closure of the Hopf hierarchy or on the linearity of the underlying transport problem. Since the hidden-symmetry framework has previously been formulated for nonlinear shell models~\cite{mailybaev2022shell}, continuous transport~\cite{calascibetta2025hidden}, and Navier--Stokes turbulence~\cite{mailybaev2022hiddenMT,magacho2025scale}, we expect that the present perturbative methodology may also extend beyond the linear shell models considered here.
More broadly, the results support the view that intermittency admits a predictive perturbative description rooted in the multiplicative-cascade structure of turbulence.

\bigskip
\noindent\textit{Acknowledgments.}
We thank Luca Biferale and Massimo Cencini for multiple encouraging discussions.
A.A.M. acknowledges the hospitality of INPHYNI  during his research visits. This work was also supported by the CNPq grant 308721/2021-7, by the CAPES MATH-AmSud project CHA2MAN and by Emergence@Physique 2025.

\bigskip
\noindent\textit{Data availability.}
The scripts used to generate the figures and numerical results in this work are publicly available at \url{https://doi.org/10.5281/zenodo.20347655}.

\bibliographystyle{unsrt}

\bibliography{biblio}

\clearpage
\onecolumngrid

\setcounter{page}{1}
\setcounter{equation}{0}
\setcounter{figure}{0}
\setcounter{table}{0}
\setcounter{section}{0}

\renewcommand{\theequation}{S\arabic{equation}}
\renewcommand{\thefigure}{S\arabic{figure}}
\renewcommand{\thetable}{S\arabic{table}}

\begin{center}
{\Large \textbf{Supplemental Material}}

\vspace{0.3cm}

{\large \bf Perturbative anomalous exponents from Kolmogorov multipliers}

\vspace{0.2cm}

A. A. Mailybaev and S. Thalabard
\end{center}

\vspace{0.5cm}

This Supplemental Material presents the detailed analytical derivations and numerical procedures underlying the perturbative theory developed in the Letter for anomalous scaling in random shell models for passive scalars.
Section~I derives the stochastic equations for the multipliers and their perturbative decomposition. Section~II analyzes the zeroth-order Ornstein--Uhlenbeck process and the stationary Gaussian distribution. Section~III develops the perturbative expansion of the stationary Fokker--Planck equation and the first-order corrections to the probability density. Section~IV derives the marginal multiplier distributions and their non-Gaussian corrections. Section~V computes the anomalous scaling exponents of the structure functions. Finally, Section~VI describes the numerical simulations and statistical procedures.

\section{It\^o representation of the inertial-range dynamics}

Dynamics in the inertial interval is governed by the Stratonovich SDE system
\begin{equation}\label{eq:strat-theta-appSM}
d\theta_n
=
\Bigl(
\gamma^{2n-2}\theta_{n-1}
-\gamma^{2n}\theta_{n+1}
\Bigr)\,dt
+\varepsilon 
\Bigl(
\gamma^{n-1}\theta_{n-1}\circ dw_{n-1}
-\gamma^n \theta_{n+1}\circ dw_n
\Bigr),
\end{equation}
where $\gamma=\lambda^{1/3}$ and $w_n(t)$ are independent Wiener processes. 
The It\^o formulation of Eq.~(\ref{eq:strat-theta-appSM}) reads
\begin{equation}\label{eq:ito-theta-app}
d\theta_n
=
\Big[
\gamma^{2n-2}\theta_{n-1}
-\gamma^{2n}\theta_{n+1}
-\frac{\varepsilon^2}{2}\bigl(\gamma^{2n-2}+\gamma^{2n}\bigr)\theta_n
\Big]dt
+\varepsilon
\bigl(
\gamma^{n-1}\theta_{n-1}\,dw_{n-1}
-\gamma^n\theta_{n+1}\,dw_n
\bigr).
\end{equation}
Applying It\^o's formula, we obtain the closed It\^o system for the multipliers $x_n = \theta_n/\theta_{n-1}$ as
\begin{equation}\label{eq:xn-full-app}
\begin{aligned}
dx_n
={}&
\biggl[
\gamma^{2n-2}(1+x_n^2)
-\gamma^{2n}x_n x_{n+1}
-\gamma^{2n-4}\frac{x_n}{x_{n-1}}
\biggr]dt
+\varepsilon^2 x_n
\biggl[
\gamma^{2n-4}\Bigl(\frac12+\frac{1}{x_{n-1}^2}\Bigr)
+\gamma^{2n-2}(1+x_n^2)
-\frac{\gamma^{2n}}{2}
\biggr]dt
\\
&\quad
+\varepsilon
\biggl[
-\gamma^{n-2}\frac{x_n}{x_{n-1}}\,dw_{n-2}
+\gamma^{n-1}(1+x_n^2)\,dw_{n-1}
-\gamma^n x_n x_{n+1}\,dw_n
\biggr].
\end{aligned}
\end{equation}
The drift term contains contributions inherited from the deterministic dynamics as well as the It\^o correction, while the noise is local in scale and involves three neighboring Wiener processes \(dw_{n-2},dw_{n-1},dw_n\).
Representing $x_n=\gamma^{-1}+\varepsilon z_n$, we obtain the transformed SDE system
\begin{equation}
dz_n = a_n(z,\varepsilon)\,dt+\sum_{m} B_{nm}(z,\varepsilon)\,dw_m,
\label{eq_SDE_zSM}
\end{equation}
with the drift 
\begin{equation}
\begin{aligned}
a_n(z,\varepsilon)
={}&
\gamma^{2n-3}
\Bigg[
2z_n-\gamma^2(z_n+z_{n+1})
-\frac{z_n-z_{n-1}}{1+\varepsilon\gamma z_{n-1}}
+\varepsilon\gamma\bigl(z_n^2-\gamma^2 z_n z_{n+1}\bigr)
\Bigg]
\\
&\quad
+\varepsilon\,\gamma^{2n-5}(1+\varepsilon\gamma z_n)
\Bigg[
\frac12+\frac{\gamma^2}{(1+\varepsilon\gamma z_{n-1})^2}
+\gamma^2+(1+\varepsilon\gamma z_n)^2-\frac{\gamma^4}{2}
\Bigg]
\end{aligned}
\label{eq:drift_an}
\end{equation}
and noise coefficients
\begin{equation}
\begin{aligned}
B_{nm}(z,\varepsilon)
=
-\gamma^{n-2}\frac{1+\varepsilon\gamma z_n}{1+\varepsilon\gamma z_{n-1}}\,\delta_{m,n-2}
+\gamma^{n-3}\Bigl[\gamma^2+(1+\varepsilon\gamma z_n)^2\Bigr]\delta_{m,n-1}
-\gamma^{n-2}(1+\varepsilon\gamma z_n)(1+\varepsilon\gamma z_{n+1})\,\delta_{m,n}.
\end{aligned}
\label{eq:diffusion_Bnm}
\end{equation}

System \eqref{eq_SDE_zSM} possesses a hidden scaling symmetry:
\begin{equation}
z_n(t) \mapsto z_{n+s}(\gamma^{-2s}t),
\quad
w_n(t) \mapsto \gamma^s\,w_{n+s}(\gamma^{-2s}t),
\quad
s \in \mathbb{Z}.
\label{eq_HSSM}
\end{equation}
The assumption that the hidden scaling symmetry is restored statistically plays a central role in the perturbative construction developed below. In particular, it implies that the stationary statistics in the inertial interval are invariant under shell translations, which leads naturally to the Toeplitz structure of the covariance matrix and to the translation-invariant structure of higher-order correction tensors. In this sense, the inertial-range dynamics can be viewed as a statistically homogeneous system in logarithmic scale space.

The drift vector \(a(z,\varepsilon)\) and the noise matrix \(B(z,\varepsilon)\) admit the expansions
\begin{align}
a(z,\varepsilon) &= A^{(0)} z+\varepsilon a^{(1)}(z)+\varepsilon^2 a^{(2)}(z)+\cdots,
\label{eq_a_exp} \\
B(z,\varepsilon) &= B^{(0)}+\varepsilon B^{(1)}(z)+\varepsilon^2 B^{(2)}(z)+\cdots,
\label{eq_B_exp}
\end{align}
where \(A^{(0)}\) and \(B^{(0)}\) are constant matrices. The zeroth- and first-order terms are given by
\begin{align}
a_n^{(0)}(z)
&= (A^{(0)} z)_n
= \gamma^{2n-3}\Bigl(
z_{n-1}+(1-\gamma^2)z_n-\gamma^2 z_{n+1}
\Bigr),
\label{eq_terms_a0}
\\[2pt]
a_n^{(1)}(z)
&=
\gamma^{2n-2}\Bigl(
z_n^2+z_n z_{n-1}-z_{n-1}^2-\gamma^2 z_n z_{n+1}
\Bigr)
+\frac12\,\gamma^{2n-5}\bigl(3+4\gamma^2-\gamma^4\bigr),
\label{eq_terms_a1}
\\[2pt]
B_{nm}^{(0)}
&=
-\gamma^{n-2}\,\delta_{m,n-2}
+\bigl(\gamma^{n-1}+\gamma^{n-3}\bigr)\,\delta_{m,n-1}
-\gamma^{n-2}\,\delta_{m,n},
\label{eq_terms_B0}
\\[2pt]
B_{nm}^{(1)}(z)
&=
-\gamma^{n-1}(z_n-z_{n-1})\,\delta_{m,n-2}
+2\gamma^{n-2}z_n\,\delta_{m,n-1}
-\gamma^{n-1}(z_n+z_{n+1})\,\delta_{m,n}.
\label{eq_terms_B1}
\end{align}
We will also need the diffusion matrix
\begin{equation}
D(z,\varepsilon)= B(z,\varepsilon) B(z,\varepsilon)^T = D^{(0)}+\varepsilon D^{(1)}(z)+\varepsilon^2 D^{(2)}(z)+\cdots,
\label{eq_D_exp}
\end{equation}
whose first two terms are
\begin{equation}
D^{(0)}_{nm}
=
\gamma^{2n-6}\delta_{m,n-2}
-2(1+\gamma^2)\gamma^{2n-6}\delta_{m,n-1}
+\gamma^{2n-6}(1+4\gamma^2+\gamma^4)\delta_{m,n}
-2(1+\gamma^2)\gamma^{2n-4}\delta_{m,n+1}
+\gamma^{2n-2}\delta_{m,n+2},
\label{eq_terms_D0}
\end{equation}
\begin{align}
D^{(1)}_{nm}(z)
&=
\gamma^{2n-5}(z_{n-2}+z_n)\,\delta_{m,n-2}
-2\gamma^{2n-5}\Bigl[z_{n-1}+(2+\gamma^2)z_n\Bigr]\delta_{m,n-1}
\nonumber\\
&\quad
+\Bigl[
-2\gamma^{2n-3}z_{n-1}
+4(1+2\gamma^2)\gamma^{2n-5}z_n
+2\gamma^{2n-3}z_{n+1}
\Bigr]\delta_{m,n}
\nonumber\\
&\quad
-2\gamma^{2n-3}\Bigl[z_n+(2+\gamma^2)z_{n+1}\Bigr]\delta_{m,n+1}
+\gamma^{2n-1}(z_n+z_{n+2})\,\delta_{m,n+2}.
\label{eq_terms_D1}
\end{align}
Both matrices \(D^{(0)}\) and \(D^{(1)}(z)\) are symmetric and have bandwidth two.

\section{Zero-order Gaussian solution}

By setting $\varepsilon = 0$ in Eq.~(\ref{eq_SDE_zSM}), one obtains the Ornstein--Uhlenbeck (OU) process
\begin{equation}
dz_n = \sum_m A_{nm}^{(0)} z_m\,dt + \sum_m B_{nm}^{(0)}\,dw_m,
\label{eq_OUP}
\end{equation}
which has the Gaussian stationary density
\begin{equation}
\label{eq:GaussianSM}
p_0(z)=Z^{-1}\exp\!\left(-\tfrac12 z^T C^{-1} z\right).
\end{equation}
The covariance matrix $C$ is determined by the Lyapunov equation
\begin{equation}
\label{eq_for_C_infSM}
A^{(0)}C+CA^{(0)T}+D^{(0)}=0, \quad D^{(0)}=B^{(0)}B^{(0)T}.
\end{equation}

The assumption of statistically restored hidden symmetry (\ref{eq_HSSM}) implies that the stationary distribution (\ref{eq:GaussianSM}) is invariant under shell translations \(n \mapsto n+s\). Consequently, the covariance matrix \(C\) must be symmetric and Toeplitz,
\begin{equation}
\label{eq_C_to_c}
C_{nm} = c_{|n-m|}.
\end{equation}
We now substitute this ansatz into \eqref{eq_for_C_infSM} with the matrices (\ref{eq_terms_a0}) and (\ref{eq_terms_D0}). Evaluating the resulting equations along each diagonal \(l \ge 0\), we obtain the recurrence relations
\begin{align}
&\text{for } l=0:\qquad
2(1-\gamma^2)(c_0+c_1)+\gamma^{-3}(\gamma^4+4\gamma^2+1)=0,
\label{eq:c_l0}
\\[2mm]
&\text{for } l=1:\qquad
(1-\gamma^4)(c_1+c_2)-2(\gamma+\gamma^{-1})=0,
\label{eq:c_l1}
\\[2mm]
&\text{for } l=2:\qquad
(1-\gamma^6)c_3+(1-\gamma^2)(1+\gamma^4)c_2+(\gamma^4-\gamma^2)c_1+\gamma=0,
\label{eq:c_l2}
\\[2mm]
&\text{for } l\ge 3:\qquad
(1-\gamma^{2l+2})\,c_{l+1}
+(1-\gamma^2)(1+\gamma^{2l})\,c_l
+(\gamma^{2l}-\gamma^2)\,c_{l-1}
=0.
\label{eq:c_l3}
\end{align}

For large \(l\) and \(\gamma>1\), the dominant terms in \eqref{eq:c_l3} after multiplication by $-\gamma^{-2l}$ yield the asymptotic relation
\begin{equation}
\gamma^2 c_{l+1}+(\gamma^2-1)c_l-c_{l-1}=0.
\end{equation}
The corresponding characteristic equation \(\gamma^2 \rho^2+(\gamma^2-1)\rho-1=0\) has roots \(\rho_1=\gamma^{-2}\) and \(\rho_2=-1\).
Hence, the general asymptotic solution is
\begin{equation}
c_l \sim a_1\gamma^{-2l}+a_2(-1)^l, \qquad l\to\infty.
\end{equation}
Imposing decay of correlations selects the branch \(c_l \propto\gamma^{-2l}\), which determines a unique admissible solution. 

\begin{figure}[tp]
\centering
\includegraphics[width=0.5\textwidth]{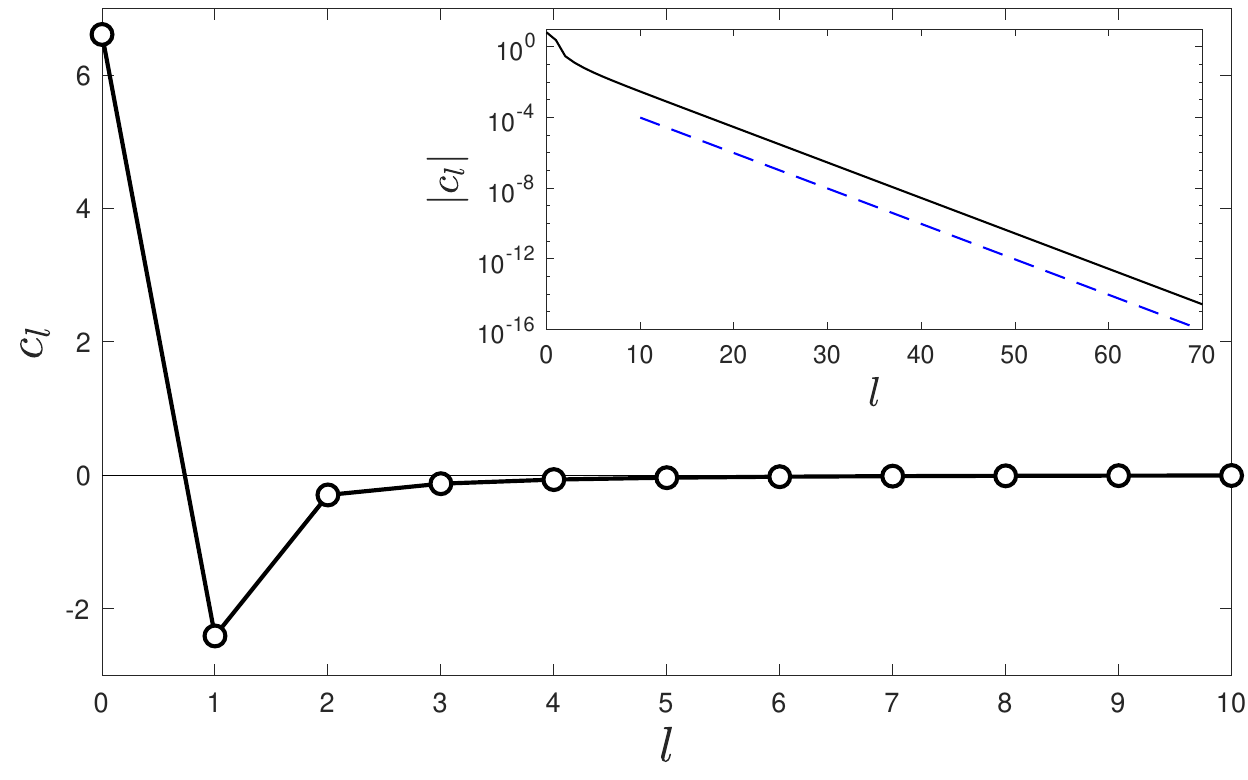}
\caption{Covariance coefficients $c_l$ for $\gamma=2^{1/3}$. Inset: $|c_l|$ in vertical log-scale for the numerical solution with cutoff $l_{\max}=70$. The dashed line shows the asymptotics $\propto \gamma^{-2l}$.}
\label{figSM1}
\end{figure}

Given $\gamma$, the solution can be obtained numerically by introducing a cutoff $l_{\max}$ and setting $c_l=0$ for $l>l_{\max}$. Then the linear system~(\ref{eq:c_l0})--(\ref{eq:c_l3}) with $l=0,1,\ldots,l_{\max}$ can be solved with respect to $c_0,\ldots,c_{l_{\max}}$. This procedure converges as $l_{\max}$ increases, providing a stable and efficient computation of the covariance coefficients; see Fig.~\ref{figSM1}.

It is useful to derive the expression for the sum of covariance coefficients. 
Let us multiply each Eq.~(\ref{eq:c_l3}) by $\gamma^{-2l}$ and sum over $l \ge 3$. Taking into account that $c_l \propto \gamma^{-2l} \to 0$ as $l \to \infty$ yields 
\begin{equation}
\sum_{j \ge 4} c_j
= -\frac{(1-\gamma^{-4})c_2
+(2-\gamma^2-\gamma^{-4})c_3}{2(1-\gamma^2)}.
\end{equation}
Adding to this expression $c_1+c_2+c_3$ and using Eqs.~(\ref{eq:c_l0})--(\ref{eq:c_l2}), we find
\begin{equation}
\sum_{l\ge 1} c_{l}
=
-\frac{c_0}{2}+\frac{\gamma^2+1}{4\gamma^3}.
\label{eq_sum_cl}
\end{equation}

\section{Perturbation expansion of the stationary Fokker--Planck equation}

The stationary Fokker--Planck (FP) equation for system \eqref{eq_SDE_zSM} has the form
\begin{equation}
-\sum_n \partial_n \bigl(a_n(z,\varepsilon)\,p\bigr) 
+\frac12\sum_{n,m} \partial_n\partial_m\bigl(D_{nm}(z,\varepsilon)\,p\bigr) = 0.
\label{eq:FP_stationary}
\end{equation}
The drift and diffusion coefficients admit the perturbative expansions
\eqref{eq_a_exp} and \eqref{eq_D_exp}. Accordingly, we seek a stationary
solution in the form of a perturbation series
\begin{equation}
p(z,\varepsilon)=p_0(z)\bigl[1+\varepsilon r_1(z)+\varepsilon^2 r_2(z)+\cdots\bigr],
\label{eq_p_expansion}
\end{equation}
with normalization conditions
\begin{equation}
\int r_k(z)p_0(z)\,dz=0, \quad k \ge 1.
\label{eq_NC}
\end{equation}
Substituting these expansions into Eq.~\eqref{eq:FP_stationary} and collecting
powers of \(\varepsilon\), we obtain a hierarchy of equations. 

At order \(\varepsilon^0\), we recover
\begin{equation}
-\sum_n \partial_n\Bigl( \big(A^{(0)}z\big)_n p_0\Bigr)
+\frac12\sum_{n,m} D^{(0)}_{nm}\,\partial_n\partial_m p_0 = 0,
\label{eq_perut_order0}
\end{equation}
which represents the FP equation for the OU process (\ref{eq_OUP}) with the Gaussian solution (\ref{eq:GaussianSM}).

At order $\varepsilon^1$, after taking the derivatives and cancelling the common factor $p_0(z)$, one obtains
\begin{equation}
\begin{aligned}
&
-\sum_n\bigl[(A^{(0)}+D^{(0)}Q)z\bigr]_n\partial_n r_1
+\frac12\sum_{n,m}D^{(0)}_{nm}\partial_n\partial_m r_1
=
\sum_n\bigl[\partial_n a_n^{(1)}-(Qz)_n a_n^{(1)}\bigr]
\\
&
-\frac12\sum_{n,m}
\Bigl[
\partial_n\partial_m D^{(1)}_{nm}
-(Qz)_m\,\partial_n D^{(1)}_{nm}
-(Qz)_n\,\partial_m D^{(1)}_{nm}
+\bigl((Qz)_n(Qz)_m-Q_{nm}\bigr)D^{(1)}_{nm}
\Bigr], \quad Q = C^{-1}.
\end{aligned}
\label{eq:M0_operator_r}
\end{equation}
Similarly, higher orders $\varepsilon^k$, $k \ge 2$, yield a hierarchy of equations for correction terms $r_k$.

We remark that the perturbative expansion developed here should be understood as an asymptotic expansion around a scale-invariant Gaussian fixed point associated with the hidden symmetry (\ref{eq_HSSM}). In particular, it does not capture exponentially small non-perturbative contributions, such as those associated with sign changes of multipliers. Within this picture, the non-Gaussian corrections generated by the perturbation series describe deviations from the fixed-point measure while preserving the statistically homogeneous structure of the inertial-range dynamics in logarithmic scale space.

\subsection{Cubic terms of the first-order correction}

Using the explicit expressions \eqref{eq_terms_a1} and \eqref{eq_terms_D1}, one
sees that the right-hand side of \eqref{eq:M0_operator_r} contains only terms
that are linear or cubic in \(z_n\). The structure of the left-hand side then
implies that the unknown function \(r_1(z)\) must also be a polynomial containing
only linear and cubic terms. The normalization condition
\eqref{eq_NC} is then satisfied automatically, since these terms are odd
with respect to \(z \mapsto -z\). 

The derivation of the cubic correction proceeds in several steps. First, we rewrite the first-order Fokker--Planck equation in terms of the dual Gaussian variables \(q=C^{-1}z\), which diagonalize Gaussian contractions. We then exploit translation and permutation symmetries to reduce the cubic coefficients to a two-index tensor \(W_{xy}\). Finally, substituting the polynomial ansatz into the perturbative equation yields a closed linear system for \(W\).

Let \(r_1^{(c)}(z)\) denote the cubic part of the unknown function \(r_1(z)\).
Extracting the cubic terms in \eqref{eq:M0_operator_r} and using the identity
\[
A^{(0)}+D^{(0)}C^{-1}=-CA^{(0)T}C^{-1},
\]
which follows from \eqref{eq_for_C_infSM}, we obtain the equation for \(r_1^{(c)}(z)\) in the form
\begin{equation}
\sum_n \Bigl(CA^{(0)T}C^{-1}z\Bigr)_n \partial_n r_1^{(c)} 
= -\sum_n (C^{-1}z)_n \alpha_n(z)
-\frac12\sum_{n,m}
(C^{-1}z)_n(C^{-1}z)_mD^{(1)}_{nm}(z),
\label{eq:r_cubic}
\end{equation}
where
\begin{equation}
\label{eq_alpha_n}
\alpha_n(z)
= \gamma^{2n-2}\bigl(
z_n^2+z_n z_{n-1}-z_{n-1}^2-\gamma^2 z_n z_{n+1}
\bigr)
\end{equation}
is the quadratic part of \(a_n^{(1)}(z)\) from Eq.~\eqref{eq_terms_a1}.

It is convenient to solve this equation using the dual Gaussian variables
\(q=(q_n)_{n}\), defined by
\begin{equation}
\label{eq_dual_G}
z = Cq, \quad q = C^{-1}z, \quad \nabla_q = C \nabla_z,
\end{equation}
where the last identity gives the relation between derivatives.
In these variables, \eqref{eq:r_cubic} reads
\begin{equation}
\label{eq_FO_cubic_C}
\sum_n \big(A^{(0)T}q\big)_n \,\frac{\partial r_1^{(c)}}{\partial q_n}
=
-\sum_n q_n \alpha_n(Cq)
-\frac12\sum_{n,m}q_n q_m D^{(1)}_{nm}(Cq).
\end{equation}
We seek a cubic solution in the form
\begin{equation}
r_1^{(c)} = \sum_{i,j,k} w_{ijk} \,q_i q_j q_k.
\label{eq:r1_cubic_ansatz_app}
\end{equation}
Since the coefficients \(w_{ijk}\) are symmetric and translation invariant in the indices, we write
\begin{equation}
w_{ijk}=W_{xy},\quad x=j-i,\quad y=k-j.
\label{eq:W_xy_definition}
\end{equation}
Permutation symmetry implies
\begin{equation}
W_{xy}
=
W_{-x,\,x+y}
=
W_{x+y,\,-y}
=
W_{-y,\,-x}
=
W_{y,\,-x-y}
=
W_{-x-y,\,x}.
\label{eq:W_xy_symmetry}
\end{equation}
Therefore, it is sufficient to solve the problem in the fundamental sector
\begin{equation}
x\ge 0,\quad y\ge 0,
\label{eq:fund_sector_app}
\end{equation}
and recover the values outside this sector using the symmetry relations \eqref{eq:W_xy_symmetry}.

Substituting \eqref{eq_terms_a0}, \eqref{eq_terms_D1}, \eqref{eq_C_to_c},
\eqref{eq_alpha_n}, \eqref{eq:r1_cubic_ansatz_app}, and \eqref{eq:W_xy_definition} into (\ref{eq_FO_cubic_C}),
and performing lengthy but straightforward manipulations, we obtain in the fundamental sector the system
\begin{equation}
\label{eq_W_sys}
W_{x+1,y}
+(1-\gamma^2)W_{xy}
-\gamma^2 W_{x-1,y} = \frac{F_{xy}}{3},
\qquad x \ge 0,\ y \ge 0,
\end{equation}
where 
\begin{align}
F_{xy}
=\,
& \gamma c_{x+1}c_{x+y+1}
+\gamma^3 c_x c_{|x+y-1|}
-\gamma c_x c_{x+y}
-\gamma c_x c_{x+y+1}
-\frac{\gamma^2}{2}\,\delta_{x,2}\bigl(c_{y+2}+c_y\bigr)
\nonumber\\
&+\delta_{x,0}\bigl[c_{y+1}-2(1+2\gamma^2)\gamma^{-2}c_y-c_{|y-1|}\bigr]
+\delta_{x,1}\bigl[c_{y+1}+(2+\gamma^2)c_y\bigr].
\label{eq:F_sector_semiexplicit}
\end{align}
The terms with \(x=-1\) in Eq.~(\ref{eq_W_sys}) must be expressed through the symmetry relations as
\begin{equation}
\label{eq_W_sym_xneg}
W_{-1,0} = W_{0,1}, \qquad
W_{-1,y} = W_{1,y-1}, \quad y \ge 1.
\end{equation}
Relations \eqref{eq_W_sys}--\eqref{eq_W_sym_xneg} define the linear problem for
\(W_{xy}\) in the fundamental sector \eqref{eq:fund_sector_app}. Once \(W\) is found,
the cubic part of the first-order correction is reconstructed from
\eqref{eq:r1_cubic_ansatz_app}--\eqref{eq:W_xy_symmetry}.

\begin{figure}[tp]
\centering
\includegraphics[width=0.7\textwidth]{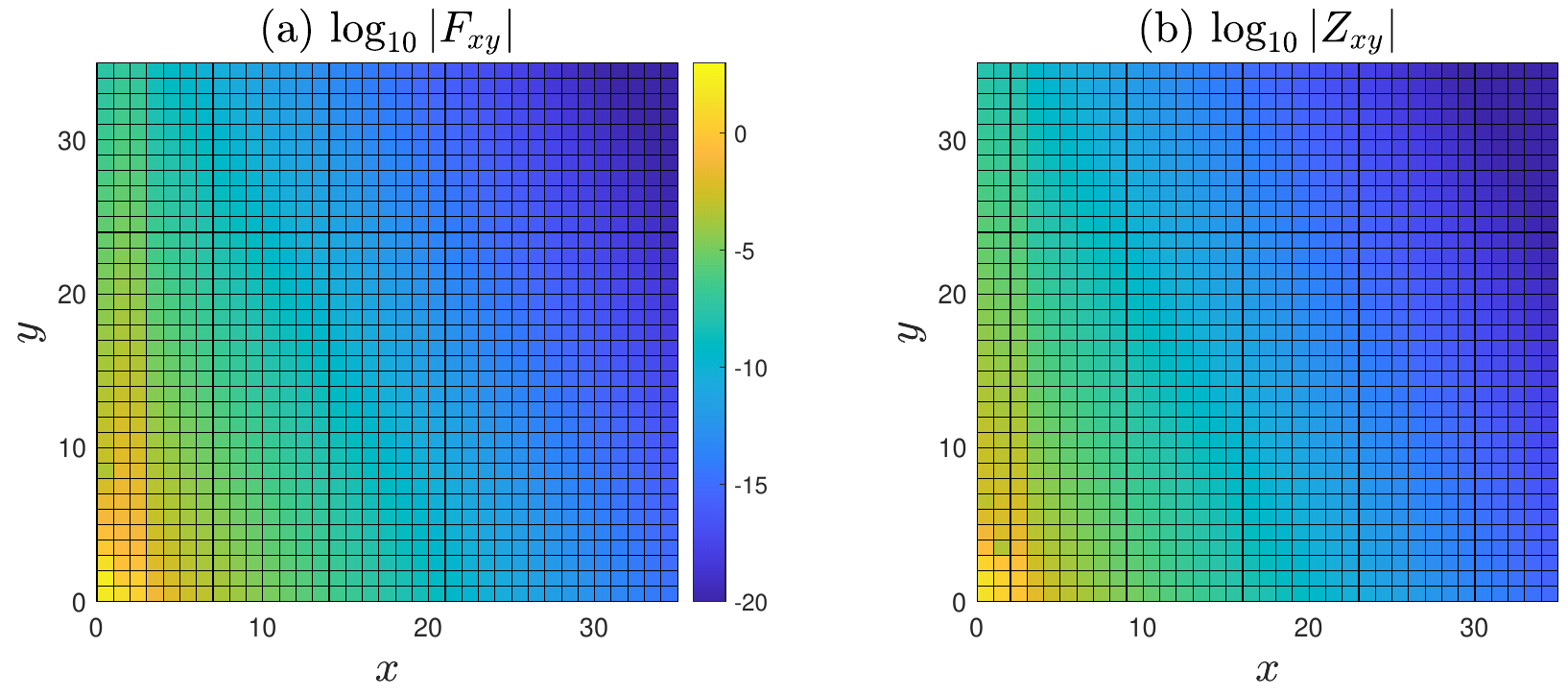}
\caption{Color plots of (a) $\log_{10}|F_{xy}|$ and (b) $\log_{10}|Z_{xy}|$, showing the matrix elements that determine the non-Gaussian corrections. The computations are performed for $\gamma=2^{1/3}$ with cutoffs $x_{\max}=y_{\max}=35$.}
\label{figSM2}
\end{figure}

Given \(\gamma\) and the covariance coefficients \(c_l\), the matrix \(W\) can be computed numerically by solving the linear system \eqref{eq_W_sys}--\eqref{eq_W_sym_xneg} with cutoffs \(x\le x_{\max}\) and \(y\le y_{\max}\) imposed at sufficiently large values. The solution converges as the cutoffs increase. In the case \(\gamma=2^{1/3}\), we obtain \(W_{00}\approx 25.8962\).
We remark that it is useful to express the solution in terms of the variables
\begin{equation}
\label{eq:Z_from_W}
Z_{xy}=
\begin{cases}
W_{x-1,0}+W_{x0}, & x\ge 1,\ y = 0,\\[1mm]
W_{0y}-W_{0,y-1}, & x=0,\ y \ge 1,\\[1mm]
W_{x-1,y}+W_{xy}-W_{x-1,y-1}-W_{x,y-1}, & x\ge 1,\ y \ge 1,
\end{cases}
\end{equation}
with \(Z_{00}=W_{00}\), in which case both $F_{xy} \to 0$ and $Z_{xy} \to 0$ as $|x|+|y| \to \infty$; see Fig.~\ref{figSM2}.

\subsection{Linear terms of the first-order correction}

It is convenient to express the perturbative correction in Wick-ordered form. This representation isolates the genuinely non-Gaussian part of the perturbation by removing lower-order Gaussian contractions, while naturally separating the mean shift from the irreducible cubic correction. As a result, the perturbative structure becomes particularly transparent in terms of the dual Gaussian variables.

To complete the first-order correction, it is convenient to combine the linear and cubic contributions in the Wick-ordered form
\begin{equation}
r_1(z) =  m \sum_n q_n+\sum_{i,j,k} w_{ijk}\, {:}\, q_i q_j q_k\,{:}\,,
\label{eq:r_allSM}
\end{equation}
where \(q=C^{-1}z\) are the dual Gaussian variables and
\begin{equation}
{:}\, q_i q_j q_k \,{:}
=
q_i q_j q_k
-\langle q_i q_j \rangle_0 q_k
-\langle q_i q_k \rangle_0 q_j
-\langle q_j q_k \rangle_0 q_i
\end{equation}
is the Wick-ordered product. Here \(\langle\cdot\rangle_0\) denotes averaging with respect to the Gaussian density \eqref{eq:GaussianSM}, in which case \(\langle q_i q_j\rangle_0 = C^{-1}_{ij}\).
By translation invariance, the linear part is characterized by a single scalar coefficient $m$. In the Wick-ordered representation, this coefficient determines the mean shift, as see below.
Using expressions (\ref{eq:GaussianSM}) and (\ref{eq:r_allSM}) in the expansion (\ref{eq_p_expansion}), we write 
\begin{equation}
p(z,\varepsilon)=Z^{-1}
\exp\!\left(-\frac12 z^T C^{-1} z\right)
\Bigl[1+\varepsilon m \sum_n q_n+\varepsilon \sum_{i,j,k} w_{ijk}\, {:}\, q_i q_j q_k \,{:} +O(\varepsilon^2)\Bigr].
\label{eq:p_shifted_wick_2}
\end{equation}
Using the relation \(q=C^{-1}z\), this probability density can also be written in the form
\begin{equation}
p(z,\varepsilon)=Z^{-1}
\exp\!\left(-\frac12 \tilde z^T C^{-1} \tilde z\right)
\Bigl[1+\varepsilon \sum_{i,j,k} w_{ijk}\, {:}\, q_i q_j q_k \,{:}\, +O(\varepsilon^2)\Bigr],
\quad
\tilde z_n = z_n-\varepsilon m,
\label{eq:p_shifted_wick}
\end{equation}
where the linear term from \eqref{eq:p_shifted_wick_2} has been absorbed into the shift \(\tilde z_n=z_n-\varepsilon m\).

We determine \(m\) from the stationarity condition for the mean. Averaging the SDE \eqref{eq_SDE_zSM} with respect to the stationary distribution \(p(z)\), we obtain
\begin{equation}
\langle a_n(z,\varepsilon)\rangle 
= \int a_n(z,\varepsilon)p(z,\varepsilon)\, dz
= 0,
\label{eq_mean_order}
\end{equation}
where the mean is considered with respect to the perturbed density $p(z,\varepsilon)$ for a given $\varepsilon > 0$.
Substituting the expansions \eqref{eq_a_exp} and \eqref{eq:p_shifted_wick}, the terms of order $\varepsilon^1$ satisfy
\begin{equation}
m \sum_i \Big\langle \big(A^{(0)}z\big)_n q_i \Big\rangle_0
+ \sum_{i,j,k} w_{ijk} \Big\langle \big(A^{(0)}z\big)_n \,{:}\, q_i q_j q_k \,{:} \Big\rangle_0
+ \big\langle a_n^{(1)}(z)\big\rangle_0 = 0.
\label{eq_first_order}
\end{equation}
Since Wick ordering removes all Gaussian contractions of the cubic term with a linear variable, one has
\begin{equation}
\Big\langle \big(A^{(0)}z\big)_n \,{:}\, q_i q_j q_k \,{:} \Big\rangle_0 = 0.
\label{eq_Azqqq}
\end{equation}
Using the property $\big\langle z_k q_n \big\rangle_0 = \delta_{kn}$ of dual Gaussian variables, we find
\begin{equation}
\sum_i \Big\langle \big(A^{(0)}z\big)_n q_i \Big\rangle_0 
=
\sum_{i,k} A^{(0)}_{nk}\,\langle z_k q_i\rangle_0
= \sum_k A^{(0)}_{nk} = 2(1-\gamma^2)\gamma^{2n-3},
\end{equation}
where the last equality follows from Eq.~\eqref{eq_terms_a0}.
Next, using \eqref{eq_terms_a1} with the Gaussian moments \(\langle z_n^2\rangle_0=c_0\) and
\(\langle z_n z_{n\pm1}\rangle_0=c_1\), we obtain
\begin{align}
\big\langle a_n^{(1)}(z)\big\rangle_0
&=
\gamma^{2n-2}\Bigl(
\langle z_n^2\rangle_0+\langle z_n z_{n-1}\rangle_0
-\langle z_{n-1}^2\rangle_0
-\gamma^2\langle z_n z_{n+1}\rangle_0
\Bigr)
+\frac12\gamma^{2n-5}(3+4\gamma^2-\gamma^4)
\nonumber\\
&=
\gamma^{2n-2}(1-\gamma^2)c_1
+\frac12\gamma^{2n-5}(3+4\gamma^2-\gamma^4).
\label{eq:a1_mean}
\end{align}

Substituting Eqs.~\eqref{eq_Azqqq}--\eqref{eq:a1_mean} into \eqref{eq_first_order}, we obtain
\begin{equation}
m
=
-\frac{\gamma c_1}{2}
-\frac{3+4\gamma^2-\gamma^4}{4\gamma^2(1-\gamma^2)}
= \frac{\gamma}{2}\,c_0-\frac{\gamma^2+1}{2\gamma^2},
\label{eq:m_finalSM}
\end{equation}
where we used Eq.~(\ref{eq:c_l0}) in the last expression.
Thus, the first-order correction is naturally represented by a shifted Gaussian measure together with a Wick-ordered cubic perturbation (\ref{eq:p_shifted_wick}), and the shift is determined by Eq.~\eqref{eq:m_finalSM}.

\section{Marginal distributions}

The marginal distribution is defined by
\begin{equation}
p(z_n)=\int p(z)\prod_{k\ne n}dz_k,
\label{eq_p_marg}
\end{equation}
where, by a slight abuse of notation, we use the same symbol \(p\) for the full and marginal probability densities. We now compute the marginal distribution using representation \eqref{eq:p_shifted_wick_2}.
Using the centered Gaussian relations $\langle q_i z_n\rangle_0=\delta_{in}$ and $\langle z_n^2\rangle_0=c_0$,
and Wick's theorem, we find
\begin{equation}
\langle q_i | z_n\rangle_0=\frac{\delta_{in}}{c_0}\,z_n, \quad
\langle \,{:}\,q_iq_jq_k\,{:} \,| z_n\rangle_0
=
\delta_{in}\delta_{jn}\delta_{kn}
\left(
\frac{z_n^3}{c_0^3}-\frac{3z_n}{c_0^2}
\right).
\label{eq:wick_conditional_marginal}
\end{equation}
Therefore, only the diagonal coefficient \(w_{nnn}=W_{00}\) contributes, and
\begin{equation}
\sum_{i,j,k} w_{ijk}\,
\langle \,{:}\,q_iq_jq_k\,{:}\, | z_n\rangle_0
=
\frac{W_{00}}{c_0^3}
\left( z_n^3-3c_0 z_n \right).
\label{eq:rc_conditional_marginal}
\end{equation}
Substituting representation \eqref{eq:p_shifted_wick_2} into \eqref{eq_p_marg} and using expressions (\ref{eq:wick_conditional_marginal}) and (\ref{eq:rc_conditional_marginal}), we arrive at
\begin{equation}
p(z_n) = 
\frac{1}{\sqrt{2\pi c_0}}
\exp\!\left(-\frac{z_n^2}{2c_0}\right)
\left[
1+\varepsilon\frac{m}{c_0}z_n+\varepsilon \frac{W_{00}}{c_0^3}
\left(z_n^3-3c_0 z_n\right) +O(\varepsilon^2) 
\right].
\label{eq:marginal_shifted_final_simplifiedSM}
\end{equation}
This probability density can also be written as
\begin{equation}
p(z_n) = 
\frac{1}{\sqrt{2\pi c_0}}
\exp\!\left(-\frac{(z_n-\varepsilon m)^2}{2c_0}\right)
\left[
1+\varepsilon \frac{W_{00}}{c_0^3}
\left(z_n^3-3c_0 z_n\right) +O(\varepsilon^2) 
\right].
\label{eq:marginal_shifted_final_simplifiedBSM}
\end{equation}
Thus, the marginal distribution is a Gaussian with mean \(\varepsilon m\) and variance \(c_0\), multiplied by the cubic correction determined by \(W_{00}\). 

We remark that the perturbative expression (\ref{eq:marginal_shifted_final_simplifiedBSM}) is an asymptotic expansion in \(\varepsilon\). Accordingly, a finite-order truncation of the series does not necessarily define a positive probability density for sufficiently large values of \(|z_n|\). This behavior is a standard feature of asymptotic perturbative expansions and does not affect the validity of the expansion for the computation of statistical observables.

\section{Perturbation expansion of the power-law exponents}

In this section, we derive the perturbative corrections to the scaling exponents of the structure functions. We may assume that all multipliers \(x_n=\theta_n/\theta_{n-1}\) remain positive, which allows us to expand in \(\varepsilon\) the telescopic representation 
\begin{equation}
|\theta_n|^p
= \prod_{k=1}^n x_k^p
= \prod_{k=1}^n (\gamma^{-1}+\varepsilon z_k)^p
= \gamma^{-pn} \prod_{k=1}^n (1+\varepsilon \gamma z_k)^p.
\label{eq_tele_theta}
\end{equation}
Strictly speaking, sufficiently large negative fluctuations of \(z_k\) may produce sign changes of individual multipliers \(x_k=\gamma^{-1}+\varepsilon z_k\). However, in the leading Gaussian approximation for small \(\varepsilon\), the probability of such an event scales as
\begin{equation}
\mathbb{P}(x_k<0)
=
\mathbb{P}\left(z_k<-\frac{1}{\varepsilon\gamma}\right)
\propto
\exp\left(
-\frac{1}{2c_0\gamma^2\varepsilon^2}
\right),
\end{equation}
up to subleading prefactors. These non-perturbative events are therefore beyond all algebraic orders in \(\varepsilon\) and do not affect the perturbative asymptotics of the scaling exponents.

Expanding Eq.~(\ref{eq_tele_theta}) in powers of \(\varepsilon\) for an arbitrary exponent \(p \in \mathbb{R}\), we obtain
\begin{equation}
|\theta_n|^p
= 
\gamma^{-pn}
\Bigg[
1
+p\varepsilon \gamma \sum_{k=1}^n z_k 
+\frac{\varepsilon^2 \gamma^2}{2}
\left(
p^2 \Big(\sum_{k=1}^n z_k\Big)^2
- p \sum_{k=1}^n z_k^2
\right)
\Bigg]
+O(\varepsilon^3).
\label{eq_prod_tel}
\end{equation}
Taking the expectation, we write
\begin{equation}
S_p(n)
=
\gamma^{-pn}
\left[
1
+p\varepsilon \gamma \sum_{k=1}^n \langle z_k\rangle
+\frac{\varepsilon^2 \gamma^2}{2}
\left(
p^2 \left\langle \Big(\sum_{k=1}^n z_k\Big)^2 \right\rangle
- p \sum_{k=1}^n \langle z_k^2\rangle
\right)
\right]
+O(\varepsilon^3).
\label{eq:Sp_prelim}
\end{equation}

We now compute the averages using the expansion \eqref{eq:p_shifted_wick_2}.
For the Gaussian averages $\langle \cdot \rangle_0$, we have
\begin{equation}
\langle z_k q_n\rangle_0=\delta_{kn}, 
\quad
\langle z_iz_j q_n\rangle_0 = 0, 
\quad
\langle z_k \,{:}\, q_i q_j q_\ell \,{:}\, \rangle_0=0,
\quad
\langle z_iz_j \,{:}\, q_i q_j q_\ell \,{:}\, \rangle_0=0.
\end{equation}
Using these relations together with the expansion \eqref{eq:p_shifted_wick_2}, we obtain
\begin{equation}
\langle z_k\rangle=\varepsilon m+O(\varepsilon^2), \quad
\langle z_k z_l\rangle = c_{|k-l|}+O(\varepsilon^2).
\label{eq:zkzl_structure}
\end{equation}
Substituting \eqref{eq:zkzl_structure} into \eqref{eq:Sp_prelim}, we obtain
\begin{equation}
\begin{aligned}
S_p(n) &=
C_p\gamma^{-pn}
\Bigg[
1
+p\varepsilon^2 \gamma m\, n
+\frac{\varepsilon^2 \gamma^2}{2}
\left(
p(p-1)n c_0
+2p^2\sum_{l=1}^{n-1}(n-l)c_l
\right)+O(\varepsilon^3)
\Bigg] \\
&=
C_p\gamma^{-pn}
\exp\Bigg[
p\varepsilon^2 \gamma m\, n
+\frac{\varepsilon^2 \gamma^2}{2}
\left(
p(p-1)n c_0
+2p^2\sum_{l=1}^{n-1}(n-l)c_l
\right)+O(\varepsilon^3)
\Bigg].
\end{aligned}
\label{eq_Sp_1SM}
\end{equation}
The prefactor \(C_p\) originates from forcing-range contributions. Indeed, the distribution \eqref{eq:p_shifted_wick_2} describes only the inertial interval and is not valid in the forcing range. As a result, the multipliers associated with the first few shells (\(n\sim1\)) produce additional multiplicative contributions to the products in Eqs.~\eqref{eq_tele_theta} and \eqref{eq:Sp_prelim}. These contributions are independent of \(n\) in the large-\(n\) limit and therefore affect only the amplitudes of the structure functions.

For large \(n\), using the exponential decay \(c_l \propto \gamma^{-2l}\) together with Eq.~(\ref{eq_sum_cl}), we obtain the asymptotic expression
\begin{equation}
\sum_{l=1}^{n-1}(n-l)c_l
\approx
n\sum_{l \ge 1}c_l
-\sum_{l \ge 1} l\,c_l
=
n\left(-\frac{c_0}{2}+\frac{\gamma^2+1}{4\gamma^3}\right)
-\sum_{l \ge 1} l\,c_l.
\label{eq_sum_cl_coef}
\end{equation}
Substituting Eqs.~(\ref{eq:m_finalSM}) and (\ref{eq_sum_cl_coef}) into Eq.~(\ref{eq_Sp_1SM}) and absorbing all \(n\)-independent contributions into the prefactor \(\tilde{C}_p\), we obtain
\begin{equation}
S_p(n) \approx
\tilde{C}_p\gamma^{-pn}
\exp\left[
\frac{\gamma^2+1}{4\gamma}p(p-2)\,\varepsilon^2 n
+O(\varepsilon^3 n)
\right].
\label{eq_Sn_exp}
\end{equation}
Using \(k_n=\lambda^n=\gamma^{3n}\), this expression can be written equivalently as the asymptotic power-law scaling
\begin{equation}
S_p(n)\propto k_n^{-\zeta_p}
=
\gamma^{-3\zeta_p n},
\qquad
\zeta_p
=
\frac{p}{3}
-\frac{\gamma^2+1}{12\gamma\ln\gamma}\,
p(p-2)\,\varepsilon^2
+O(\varepsilon^4).
\end{equation}
Remarkably, the leading anomalous correction admits an explicit expression in terms of \(\gamma=\lambda^{1/3}\). Note that the error term in the expression for \(\zeta_p\) is written as \(O(\varepsilon^4)\), since odd-order corrections vanish due to the symmetry \(\varepsilon \mapsto -\varepsilon\) in the statistics of system~(\ref{eq:strat-theta-appSM}).

We emphasize that the perturbative expansion \eqref{eq_Sn_exp} for the structure function is understood in the asymptotic regime of small \(\varepsilon\) and large shell number \(n\), with \(\varepsilon^2 n = O(1)\). In this regime, the terms growing linearly with \(n\) determine the correction to the scaling exponents, while forcing-range contributions produce only \(n\)-independent corrections to the amplitudes of the structure functions.

\section{Numerical simulations}

In the numerical simulations, we take $\lambda = 2$ and $\gamma = 2^{1/3}$. We impose a cutoff at shell \(N\) by setting \(\theta_n \equiv 0\) for \(n > N\). Although the classical diffusion term \(-\kappa k_n^2 \theta_n\) vanishes in the limit \(\kappa \to 0\), it affects all scales and thus contaminates the inertial interval. To obtain a cleaner inertial range, we instead localize dissipation at the cutoff shell by introducing a dissipative term of the form \(-\gamma^{2N-1}\theta_N\).
The full system simulated is therefore
\begin{equation}
d\theta_n
=
\Bigl(
\gamma^{2n-2}\theta_{n-1}
-\gamma^{2n}\theta_{n+1}
\Bigr)\,dt
+\varepsilon 
\Bigl(
\gamma^{n-1}\theta_{n-1}\circ dw_{n-1}
-\gamma^n \theta_{n+1}\circ dw_n
\Bigr)
-\delta_{nN} \gamma^{2N-1}\theta_N\,dt,
\quad
n = 1,\ldots,N,
\end{equation}
with the forcing condition \(\theta_0 = 0\).

Time integration is performed using the Euler--Maruyama scheme for the conservative drift and stochastic parts, while the dissipative term is treated separately via a low-pass filter. The time step is chosen as $\Delta t = 0.02\,\gamma^{-2N}$.
Statistical quantities are computed over a time interval of length \(10^3\) (or longer), skipping an initial transient of length \(10^2\). The simulations exhibit period-two oscillations of statistical observables in the inertial interval. These oscillations are suppressed by averaging the statistics obtained for two consecutive cutoffs, \(N\) and \(N-1\). In the results presented here, we use \(N=23\).

We also note that, for small $\varepsilon>0$, the distributions of the shell variables $\theta_n$ acquire a positive (albeit exponentially small) probability density at $\theta_n=0$. For the multipliers $x_n=\theta_n/\theta_{n-1}$, this implies that the covariances $\mathrm{cov}(x_n,x_{n+j})$ diverge for $j=0$ and are defined only in the principal-value sense for $j\neq 0$. This exponentially small effect is not captured by the perturbative expansion in $\varepsilon$. Nevertheless, the associated divergences become apparent in numerical computations as $\varepsilon$ increases. In Fig.~2 of the main text, we therefore use a small value $\varepsilon=0.01$ to suppress this non-perturbative contribution.

\end{document}